\documentclass[pre,aps,twocolumn,showpacs]{revtex4}
\usepackage{graphicx,epsfig,amsmath,amsfonts}

\def\vec#1{{\bf #1}}
\def\text#1{{\mathrm #1}}
\begin{document}

\title{Goldstone modes in Lyapunov spectra of hard sphere systems}
\author{Astrid S. de Wijn}
\email{A.S.deWijn@phys.uu.nl}
\author{Henk van Beijeren}
\email{H.vanBeijeren@phys.uu.nl}
\affiliation{Institute for Theoretical Physics, Utrecht University, Leuvenlaan 4, 3584 CE, Utrecht, The Netherlands}

\begin{abstract}
\noindent In the study of chaotic behavior, Lyapunov exponents play an important part.
In this paper, we demonstrate how the Lyapunov exponents close to zero of a system of many hard spheres can be described as Goldstone modes, by using
a Boltzmann type of approach.
At low densities, the correct form is found for the wave number dependence of the exponents
as well as for
the corresponding eigenvectors in tangent-space.
The predicted values for the Lyapunov exponents belonging to the transverse mode are within a
few percent of the values found in recent simulations,
the propagation velocity for the longitudinal mode is within 1 \%,
but the value for the Lyapunov exponent belonging to the longitudinal mode deviates from the
simulations by 30\%.
For higher densities, the predicted values deviate more from the values calculated in the
simulations.
These deviations may be due to contributions from ring collisions and similar terms, which,
even at low densities, can contribute to the leading order.
\end{abstract}

\date{\today}

\maketitle

\section{Introduction}
In the past years, interest in the connections between dynamical-systems theory and statistical mechanics has increased and many important results have been obtained.
Some of the interest has been directed towards the connections between chaoticity and the decay to equilibrium.
Gallavotti and Cohen \cite{gc1,gc2} put forward a chaotic hypothesis, conjecturing that many-particle systems as studied by statistical mechanics will generically be strongly chaotic.
A central role in the study of these and related properties is played by the Lyapunov exponents, which describe the exponential separation or convergence of nearby trajectories in phase space.

Many calculations of chaotic properties have been done for variations of the Lorentz gas, which is a system of one particle bouncing between fixed spherical hard scatterers (see, for example, Refs.~\cite{long1,2d,3d,3dposch}).
Recently, calculations have also been done for the largest exponents of systems of many freely moving hard spheres \cite{ramses,prlramses}, which is a more realistic model than the single-particle system of a Lorentz gas.
Simulations for the entire spectrum of this system have been done by Posch, Hirschl, and Dellago~\cite{posch1,posch2}.
The smallest positive and corresponding negative exponents from these simulations have received a lot of attention because of their unexpected behavior.
For large enough systems, they are inversely proportional to the system length.
The tangent-space eigenvectors associated with these exponents have a wave like form.

Attempts have been made to approach these exponents by using random-matrix theory by Eckmann and Gat \cite{eckmann}, and by Taniguchi, Dettmann, and Morriss \cite{taniguchi1,taniguchi2}.
Another approach based on kinetic theory has been taken by McNamara and Mareschal \cite{mareschal}.

In this paper, we explain some of the behavior of the small exponents both qualitatively and quantitatively.
Sections \ref{sec:lyap} and \ref{sec:spheressim} are a short introduction to Lyapunov exponents and the results of the simulations done by Posch and Hirschl for hard spheres in two dimensions \cite{posch1}.
In Sec.~\ref{sec:goldstone}, we show that the small exponents are in fact due to Goldstone modes.
After an explanation of the dynamics of hard spheres in Sec.~\ref{sec:spheresdyn}, we derive a set of equations for the values of the exponents in Sec.~\ref{sec:boltzmann}.
The equations are derived by using a Boltzmann type of approach including a Sto{\ss}zahlansatz.
Finally, we discuss the general form of the solutions in Sec.~\ref{sec:solutions} and the quantitative results derived from them in Sec.~\ref{sec:results}.

\section{\label{sec:lyap}Lyapunov exponents}
Consider a system with an $\cal N$-dimensional phase space $\Gamma$.
At time $t=0$ the system is assumed to be at an initial point ${\vec\gamma}_0$ in this phase space, evolving with time according to ${\vec\gamma}({\vec\gamma}_0,t)$.
If the system is perturbed by an infinitesimal difference $\delta{\vec\gamma}_0$ in initial conditions, it evolves along  an infinitesimally different path $\gamma + \delta \gamma$, where $\delta\gamma$ is in the tangent space $\delta\Gamma$.
The evolution in tangent space is described by
\begin{eqnarray}
{\vec\gamma}({\vec\gamma}_0+\delta{\vec\gamma}_0,0)&=&{\vec\gamma}_0+\delta{\vec\gamma}_0~,\\
{\vec\gamma}({\vec\gamma}_0+\delta{\vec\gamma}_0,t)&=&{\vec\gamma}({\vec\gamma}_0,t)+\delta{\vec\gamma}({\vec\gamma}_0,t)~,\\
{\delta{\vec\gamma}({\vec\gamma}_0,t)}
\label{eq:M}&=& {{\sf M}_{{\vec\gamma}_0}(t)\cdot \delta{\vec\gamma}_0~,}
\label{eq:matrix}
\end{eqnarray}
where ${\sf M}_{{\vec\gamma}_0}(t)$ is an ${\cal N}$-dimensional matrix, defined by
\begin{eqnarray}
\label{eq:tang}
{\sf M}_{{\vec\gamma}_0}(t)=\frac{d {\vec\gamma}({\vec\gamma}_0,t)}{d {\vec\gamma}_0}~.
\end{eqnarray}
The Lyapunov exponents are the average rates of growth of such infinitesimal changes that are eigenvectors of ${\sf M}_{{\vec\gamma}_0}$,
\begin{equation}
\lambda_i =\lim_{t\rightarrow\infty} \frac{\log\mu_i(t)}{t}~,
\end{equation}
where $\mu_i(t)$ is the $i$th eigenvalue of ${\sf M}_{{\vec\gamma}_0}(t)$.
In systems which are ergodic, almost every trajectory comes infinitesimally close to any point in phase space.
This means that the Lyapunov exponents are almost independent of the initial conditions.
Often the Lyapunov exponents are defined not by using ${\sf M}_{{\vec\gamma}_0}(t)$, but $[{\sf M}_{{\vec\gamma}_0}(t)\cdot{\sf M}_{{\vec\gamma}_0}(t)^\dagger]^{1/2}$.
In the latter definition the exponents are real.
The imaginary components of the Lyapunov exponents, as we define them here, are also referred to as the winding numbers.

For a classical system of hard spheres without internal degrees of freedom, the phase space and tangent space may be represented by the positions and velocities of all particles and their infinitesimal deviations,
\begin{eqnarray}
\gamma_i & = &(\vec{r}_i, \vec{v}_i)~,\\
\delta\gamma_i & = & ({\vec{ \delta r}}_i, {\vec{ \delta v}}_i)~,
\end{eqnarray}
where $i$ runs over all particles and $\delta\gamma_i$ is the contribution of particle $i$ to $\delta\gamma$.

In the case of a purely Hamiltonian system, such as hard spheres with only the hard particle interaction, the dynamics of the system are completely invariant under time reversal.
Also, due to the incompressibility of flow in phase space, the attractor is invariant under time reversal.
Therefore, every tangent-space eigenvector that grows exponentially in forward time decreases exponentially in backward time.
Since the Lyapunov spectrum does not change under time reversal, for every positive Lyapunov exponent in such a system there is a negative exponent of equal absolute value.
This is called the conjugate pairing rule.
In systems which are reversible, but for which the attractor is not invariant under time reversal, the conditions for and the form of the conjugate pairing rule are somewhat different \cite{ramses}.

Vectors in tangent space which are generated by symmetries of the dynamics of the system do not grow or shrink exponentially.
They are eigenvectors with Lyapunov exponents 0 and are referred to as the zero modes.
For a system of hard spheres under periodic boundary conditions, these symmetries and their corresponding zero modes are uniform translations,
Galilei transformations, time translations, and velocity scaling.
They correspond to the initial displacements
\begin{eqnarray}
\delta{\vec\gamma}_i=&(\Delta\vec{r}_0,0)~,\phantom{=}&\\
\label{eq:deltav0}\delta{\vec\gamma}_i=&(0,\Delta\vec{v}_0)~,\phantom{=}&\\
\delta{\vec\gamma}_i=&(\Delta t_0\vec{v}_i,0) &= (\Delta \vec{r}_{\vec{v}},0)~,\\
\label{eq:deltavv}\delta{\vec\gamma}_i=&(0,\Delta \lambda_0\vec{v}_i)&= (0,\Delta\vec{v}_{\vec{v}})~,
\end{eqnarray}
where $\Delta\vec{r}_0$, $\Delta\vec{v}_0$, $\Delta t_0$, and $\Delta \lambda_0$ are constant vectors and scalars which are independent of the particle.
The quantities $\Delta \vec{r}_0$ and $\Delta \vec{v}_0$ can have components in all $d$ directions of the space.
In the case of Galilei transformations and velocity scaling the tangent-space vectors grow linearly, rather than exponentially, with time.  Hence the corresponding Lyapunov exponents are zero.

\section{\label{sec:spheressim}Lyapunov spectrum of hard spheres}
\begin{figure}
\includegraphics[height=8.6cm,angle=270]{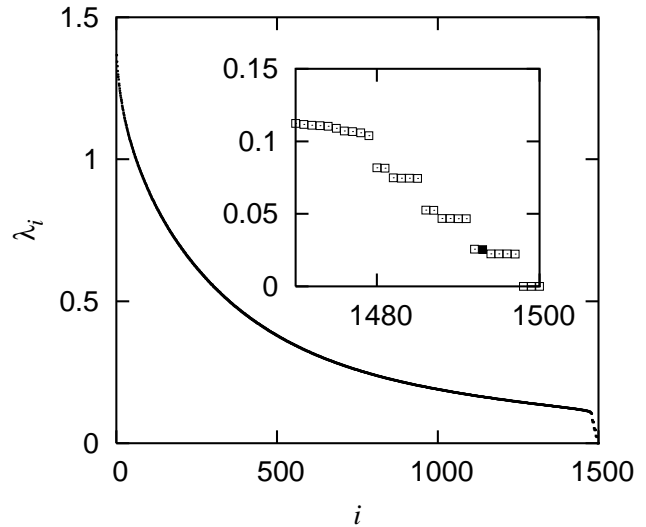}
\caption{\label{fig:simsspectrum} The spectrum of Lyapunov exponents from the simulations \cite{posch1,posch2,christinapriv} of 750 hard spheres in two dimensions at density $n=0.1$ in a rectangular box of dimensions $10 \times 75~a^2/n$, with periodic boundary conditions.
Only the positive exponents are plotted, since, by the conjugate pairing rule, the negative spectrum is exactly the opposite.
The inset shows an enlargement of the bottom right corner. 
}
\end{figure}

In principle, ${\sf M}_{{\vec\gamma}_0}(t)$ occurring in Eq.~(\ref{eq:M}) can be calculated numerically for finite times for any finite system and the eigenvalues can be determined.
Posch and Hirschl \cite{posch1} have done molecular dynamics simulations to determine the entire Lyapunov spectrum of systems consisting of many hard disks in rectangular boxes with periodic boundary conditions.
A spectrum as calculated by Posch and Hirschl is displayed in Fig.~\ref{fig:simsspectrum}.
The eigenvectors in tangent space belonging to the large exponents are typically very localized;
only a few particles closely together contribute significantly to a given eigenvector.

\begin{figure}
\includegraphics[height=8.6cm,angle=270]{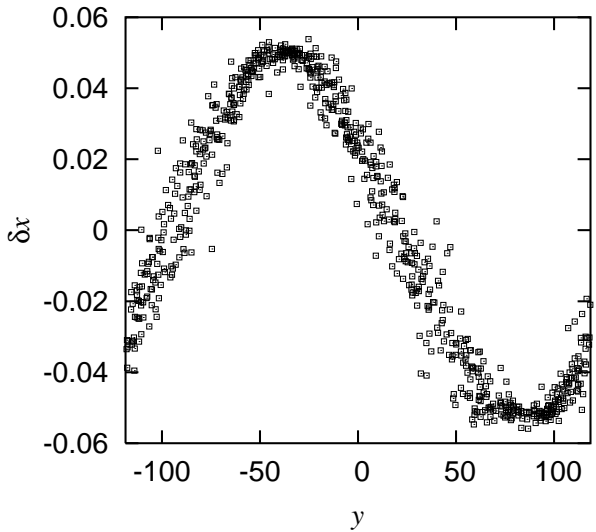}
\caption{\label{fig:simsmode}
The component of $\delta\vec{r}$ in the short direction $\delta x$ is plotted against the position of the particle in the long direction $y$ for $\lambda_{1493}$, one of the transverse modes in the first step.
These data are from the same simulation as the data in Fig.~\ref{fig:simsspectrum}.
The corresponding exponent is indicated there with a full box.}
\end{figure}

When the system is large enough compared to the mean free path, a step structure appears in the Lyapunov exponents near zero.
The size of the steps is inversely proportional to the largest dimension of the box.
The tangent-space eigenvector is distributed over all particles, much in the same way as with the zero modes.
An example of this is shown in Fig.~\ref{fig:simsmode}.

The tangent-space vectors belonging to the six exponents in each step appear, on average and to first approximation, to be linear combinations of the zero modes with a sinusoidal modulation.
This is very apparent in the example in Fig.~\ref{fig:simsmode}.
The slow modes belonging to a certain wave vector can be separated into
two groups, one consisting of four longitudinal modes and the other one of two transverse modes.
The transverse modes are found to be linear combinations of sinusoidal modulations of the zero modes resulting from a translation or a Galilei transformation in the direction perpendicular to the wave vector.
The longitudinal modes are linear combinations of modulations of the four remaining zero modes.
The transverse modes are non propagating, but the longitudinal modes propagate through the system.
This behavior has been confirmed in direct simulation Monte Carlo simulations, performed by Forster and Posch \cite{christinapriv}.
For more details on these modes, see Refs.~\cite{posch1,posch2}.

\section{\label{sec:goldstone}Goldstone Modes}
The sinusoidal modes found in the simulations may be explained as Goldstone modes.
These occur in systems with a continuous symmetry, such as the symmetries associated with the zero modes.
Translation invariance, e.g., causes the evolution operator to commute with the translation operator,
so that they have a set of common eigenfunctions.
These have the general form
\begin{equation}
\delta\gamma_i = f_\vec{k}(\vec{v}_j,\vec{r}_{ij}) \exp{(i \vec{k}\cdot\vec{r}_i)}\label{eq:goldstone}~,
\end{equation}
where
the eigenvalues of the operator translating over the vector $\vec{a}$ are of the form $\exp
i\vec{k}\cdot \vec{a}$.
The Goldstone modes are those eigenmodes that for $k\rightarrow 0$ reduce to linear combinations of the zero modes.
For nonzero values of $k$, they contain a sinusoidal modulation in space of the continuous symmetry which grows or shrinks slowly with time.
These modes were first introduced by Goldstone \cite{goldstone}.
Hydrodynamic modes and phonons in crystals are well-known examples.

In order to calculate the Lyapunov exponents belonging to these Goldstone modes, one first needs to consider the dynamics of the system.

\begin{figure}
\includegraphics[width=7cm]{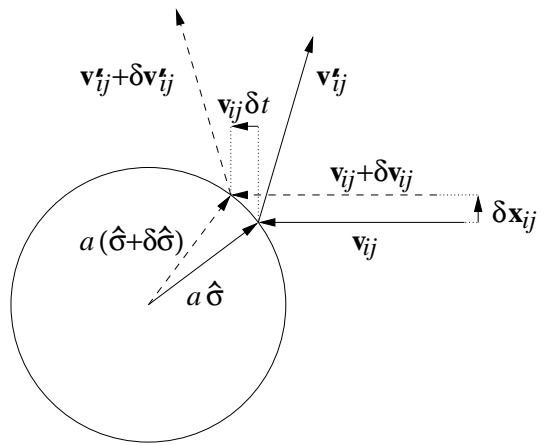}
\caption{\label{fig:bolletje} Two particles at a collision in relative phase space.  The collision normal $\hat{\vec{\sigma}}$ is the unit vector pointing from the center of one particle to the center of the other.}
\end{figure}

\section{\label{sec:spheresdyn}Dynamics of hard spheres in phase space and tangent space}
Consider a gas of identical hard spheres or disks of diameter $a$ and mass $m$ in $d$ dimensions.
The evolution in phase space consists of an alternating sequence of free flights and collisions.
During free flights the particles do not interact and the positions grow linearly with the velocities,
\begin{eqnarray}
\label{eq:freeflight1}\vec{r}_i(t) &=& \vec{r}_i(t_0) + (t - t_0)\vec{v}_i(t_0)~,\\
\vec{v}_i(t) &=& \vec{v}_i(t_0)~.
\end{eqnarray}
At a collision, momentum is exchanged between the colliding particles along the collision normal, $\hat{\vec{\sigma}} = (\vec{r}_i-\vec{r}_j)/{a}$, as shown in Fig.~{\ref{fig:bolletje}.
The other particles do not interact.  Using primes to denote the coordinates in phase space after the collision, we find
\begin{eqnarray}
\vec{r}'_i &=& \vec{r}_i~,\\
\label{eq:deltav}\vec{v}'_i &=& \vec{v}_i - \hat{\vec{\sigma}} (\hat{\vec{\sigma}} \cdot \vec{v}_{ij})~,
\end{eqnarray}
where $\vec{v}_{ij}=\vec{v}_i-\vec{v}_j$.

From Eqs.~(\ref{eq:tang}) and (\ref{eq:freeflight1}-\ref{eq:deltav}) the dynamics in tangent space can be derived \cite{ramses}.
During free flight there is no interaction between the particles and the components of the tangent-space vector transform according to
\begin{eqnarray}
\left(
\begin{array}{c}
\delta\vec{r}_i'\\
\delta\vec{v}_i'
\end{array}
\right)
&=& {\mathcal Z}(t-t_0)\cdot \left(
\begin{array}{c}
\delta\vec{r}_{i0}\\
\delta\vec{v}_{i0}
\end{array}
\right)~,\\
{\mathcal Z}(t-t_0)
&=&
\left(\begin{array}{cc}
\extracolsep{1mm}
\rule{0mm}{0mm}{\sf I}&(t-t_0){\sf I}\\
\rule{0mm}{5mm}0&{\sf I}
\end{array}\right)
~,
\label{eq:flight}
\end{eqnarray}
in which ${\sf I}$ is the $d\times d$ identity matrix.

At a collision between particles $i$ and $j$, only the contributions to the tangent-space vectors of the colliding particles are changed  \cite{prlramses}.
As shown in Fig.~\ref{fig:bolletje}, an infinitesimal difference in the positions of the particles leads to an infinitesimal change in the collision normal and collision time.
The $\vec{v} + \delta \vec{v}$ are exchanged along $\hat{\vec\sigma}+\delta\hat{\sigma}$ according to Eq.~(\ref{eq:deltav}).
This leads to infinitesimal changes in both positions and velocities right after the collision,
\begin{widetext}
\begin{eqnarray}
\left(\begin{array}{c}
\delta\vec{r}_i'\\
\delta\vec{r}_j'\\
\delta\vec{v}_i'\\
\delta\vec{v}_j'
\end{array}\right)&=&
\left(
\begin{array}{cc|cc}
\extracolsep{2mm}
\rule[-3mm]{0mm}{7mm}{\sf I-S}&{\sf S}&\rule[-5mm]{0mm}{0mm}0&0\\
\rule[-3mm]{0mm}{5mm}{\sf S}&{\sf I-S}&0&0\\
\hline
\rule[-3mm]{0mm}{7mm}{\sf -Q}&{\sf Q}&{\sf I-S}&{\sf S}\\
\rule[-3mm]{0mm}{5mm}{\sf Q}&{\sf -Q}&{\sf S}&{\sf I-S}
\end{array}\right)\cdot\left(\begin{array}{c}
\delta\vec{r}_i\\
\delta\vec{r}_j\\
\delta\vec{v}_i\\
\delta\vec{v}_j
\end{array}\right)  
\label{eq:collision}
~,
\end{eqnarray}
\end{widetext}
in which ${\sf S}$ and ${\sf Q}$ are the $d \times d$ matrices
\begin{eqnarray}
\label{eq:S}{\sf S} & = & \hat{\vec{\sigma}} \hat{\vec{\sigma}}~,\\
\label{eq:Q}{\sf Q} & = & \frac{\left[(\hat{\vec{\sigma}}\cdot\vec{v}_{ij})\,{\sf I}+\hat{\vec{\sigma}}\vec{v}_{ij}\right]\cdot
                     \left[ (\hat{\vec{\sigma}}\cdot\vec{v}_{ij})\,{\sf I}-\vec{v}_{ij} \hat{\vec{\sigma}}  \right]}
               {a (\hat{\vec{\sigma}}\cdot\vec{v}_{ij})}~.
\end{eqnarray}
Here the notation $\vec{a}\vec{b}$ denotes the standard tensor product of vectors $\vec{a}$ and $\vec{b}$.
Let ${\sf Z}(t)$ be the $N d \times N d$ matrix which performs the transformations of ${\mathcal Z}(t)$ on all particles.
Let ${\sf L}_{p}$ be the $N d \times N d$ matrix which performs the transformations of Eq.~(\ref{eq:collision}) on the two particles involved in collision $p$ and leaves the rest of the particles untouched.
${\sf M}_{\gamma_0}(t)$ in Eq.~(\ref{eq:M}) is a product of these matrices for the sequence of collisions $(1,2,\dots,p)$ between time $t$ and $t_0$.
Its specific form reads
\begin{eqnarray}
{\sf M}_{\gamma_0}(t) = {\sf Z}(t-t_{p})\cdot{\sf L}_{p}\cdot{\sf Z}(t_{p}-t_{p-1}){\sf L}_{p-1}\nonumber\\
\cdot\dots\cdot{\sf L}_{1}\cdot{\sf Z}(t_1-t_0)~.
\label{eq:product}
\end{eqnarray}

\section{\label{sec:boltzmann}Boltzmann and Enskog Equation}
Except in a calculation where the path of every particle would be calculated rigorously from the initial conditions,
as done in the molecular dynamics simulations, it is impossible to know the matrix ${\sf M}_{{\vec\gamma}_0}$ in Eq.~(\ref{eq:M}) exactly.
The tangent-space eigenvector belonging to a given Lyapunov exponent will in general
depend on the initial conditions of all the particles in a much too complicated way to specify exactly.
It is therefore impossible to know the tangent-space vector which belongs to any Lyapunov exponent exactly.
To find the exponents, one has to make some statistical approximation without allowing contributions along faster growing tangent-space vectors to blow up.
To this end, we start with assumptions similar to 
the Sto{\ss}zahlansatz in the Boltzmann equation.

\subsection{Equations in $\mu$ space}
To illustrate our calculations we first briefly review the Boltzmann and  Enskog equations
describing the dynamics of hard-sphere and hard-disk systems at low, respectively moderate densities. For
either one may start from the  first Bugoliubov-Born-Green-Kirkwood-Yvon hierarchy equation
\begin{eqnarray}
&&\frac{\partial f (\vec{r},\vec{v},t)}{\partial t} + \vec{v} \cdot \nabla_\vec{r}
f(\vec{r},\vec{v},t) + \nabla_\vec{v}\cdot \vec{a}(\vec{r},\vec{v},t) f(\vec{r},\vec{v},t)
\nonumber\\
&&\phantom{nnn}= \int_{\hat{\vec{\sigma}} \cdot (\vec{v}-\vec{u})\leq 0}\, d\vec{u}\,
d\hat{\vec{\sigma}} \,
 n a^{d-1}\, |\,\hat{\vec{\sigma}} \cdot (\vec{v}-\vec{u})|\nonumber\\
&&\phantom{nnn}\phantom{=\int} \times \left[f^{(2)}(\vec{r},\vec{v}',\vec{r}+
a\hat{\sigma},\vec{u}',t)\right.\nonumber\\
&&\phantom{nnn}\phantom{=\int\times[}\left.-f^{(2)}(\vec{r},\vec{v},\vec{r}-a\hat{\sigma},\vec{u},t)\right]~,
\label{eq:bbgky}
\label{eq:boltzmannpre}
\end{eqnarray}
in which $f^{(2)}$ denotes the  two-particle distribution function, $\vec{u}$ and $\vec{v}$ respectively
$\vec{u}'$ and $\vec{v}'$ are the velocities before the
collision with collision normal
$\hat{\vec{\sigma}}$ in the direct and restituting collisions.

The second and third term on the left-hand side of the equation, respectively, describe the effects of
free flight in position space and those of the action of external forces.
The vector $\vec{a}(\vec{r},\vec{v},t)$ is the acceleration of a particle due to external forces as a
function of position, velocity, and time.

In the low-density approximation, Boltzmann's Sto{\ss}zahlansatz approximates the precollisional pair
distribution functions in this equation by products of one-particle distributions. In addition both of
these are evaluated at the same position $\vec{r}$, which is allowed if the radius $a$ is small
compared to the mean free path. The Enskog equation is a heuristic generalization of this, known to give
a good approximate description of the dynamics
up to moderate densities (about a quarter of close-packing). In this equation the pair-distribution is
approximated by the product of two one-particle distribution functions,
evaluated at the actual positions of the two particles, and a factor $\chi_{\text E}$, equal to the
equilibrium pair correlation function at contact between the two particles evaluated as a function of
the density $n((\vec{r}_1 +\vec{r}_2)/2)$ at the point halfways between $\vec{r}_1$ and $\vec{r}_2$.
Notice that this approximation becomes exact for a system in homogeneous equilibrium. The explicit form of the Enskog
equation thus becomes

\begin{eqnarray}
\lefteqn{\frac{\partial f(\vec{r},\vec{v},t)}{\partial t} + \vec{v}\cdot\nabla_\vec{r} f(\vec{
r},\vec{v},t)} \nonumber\\
&& = \int_{\hat{\vec{\sigma}} \cdot (\vec{v}-\vec{u})\leq 0}\, d\vec{u}\, d\hat{\vec{\sigma}} \,
\chi_{\text E}(n) n a^{d-1}\, |\,\hat{\vec{\sigma}} \cdot (\vec{v}-\vec{u})|\nonumber\\
&&\phantom{=\int} \times \left[f(\vec{r},\vec{v}',t)f(\vec{r}+ a\hat{\vec{\sigma}},\vec{
u}',t)\right.\nonumber\\
&&\phantom{=\int\times}\left.-f(\vec{r},\vec{v},t)f(\vec{r}-a\hat{\vec{\sigma}},\vec{u},t)\right]~.
\label{eq:boltzmannhard}
\end{eqnarray}
This equation effectively reduces to the Boltzmann equation in the limit $n\to 0$, when the difference
in
position between the two colliding particles, $r_{ij}= a \hat{\vec{\sigma}}$, may be ignored and
$\chi_{\text E}$ approaches unity.

In
equilibrium
the time derivative term vanishes and,
in absence of external fields, the
particles are distributed homogeneously.
This yields the
Maxwellian solution
\begin{eqnarray}
\label{eq:equilibrium}f(\vec{r},\vec{v},t) = n \phi_{\text{M}}(\vec{v}) = n \left(\frac{ 2 \pi m}{k_{\text B}T}\right)^{d/2} \exp{\left( - \frac{m |\vec{v}|^2}{2 k_{\text B}T}\right)}~,
\end{eqnarray}
where $T$ is the temperature,
related to the
average kinetic energy per
particle $E$ through $E= d k_{\text B} T/2$.
In this paper, only the
equilibrium system is studied.

More details on the Boltzmann equation and Enskog's theory of dense gases
may be found in
Refs.~\cite{hcb} and \cite{chapmancowling}.

\subsection{Equations in tangent $\mu$ space}
To describe the dynamics in tangent space, a generalized Boltzmann equation must be derived for the single-particle distribution function in both $\mu$ space and "tangent $\mu$ space", $f(\vec{r},\vec{v},\delta\vec{r},\delta\vec{v},t)$.
On integration over the variables in tangent space, the equation and the solutions we are interested in must reduce to Eqs.~(\ref{eq:boltzmannhard}) and (\ref{eq:equilibrium}) respectively.

For given initial conditions the eigenvectors of ${\sf M}_{{\vec\gamma}_0}$ in general depend sensitively on the precise values of the collision parameters of all collisions, as generated by the positions and velocities of all particles.
The zero modes are exceptions to this.
For small $k$ it is to be expected that the Goldstone modes behave in a similar way and are approximately independent of the collision parameters of the various collisions.
Under those circumstances one may expect that the tangent-space vectors
$\delta\vec{r}$ and $\delta\vec{v}$ can be described by a single-particle distribution function that 
depends smoothly on velocity, position and time, just like the velocity distribution in ordinary 
$\mu$ space. If in addition one makes the assumption that the distribution function of the tangent space 
vectors of two particles about to collide, factorizes in a similar way as the distribution of their 
velocities, one ends up with a generalized Enskog equation in tangent $\mu$ space, which, in absence of an 
external field, is of the form
\begin{widetext}
\begin{eqnarray}
\lefteqn{\frac{\partial f(\vec{r},\vec{v},\delta\vec{r},\delta\vec{v},t)}{\partial t}
    + \vec{v}\cdot\nabla_\vec{r} f(\vec{r},\vec{v},\delta\vec{r},\delta\vec{v},t)
    \phantom{+}+ \delta\vec{v}\cdot\nabla_{\delta \vec{r}} f(\vec{r},\vec{v},\delta\vec{r},\delta\vec{v},t)}
\nonumber\\
&=& \int_{\hat{\vec{\sigma}} \cdot (\vec{v}-\vec{u})\leq 0}\, d\vec{u}\,d\hat{\vec{\sigma}}\, d\delta\vec{s}\, d\delta\vec{u} \, \chi_{\mathrm E}(n)n a^{d-1}\,
|\,\hat{\vec{\sigma }} \cdot (\vec{v}-\vec{u})|\nonumber\\ &&\phantom{=\int}
\times[f(\vec{r},\vec{v}',\delta\vec{r}',\delta\vec{v}',t)f(\vec{r}+a\hat{\vec{\sigma}},\vec{u}',\delta\vec{s}',\delta\vec{u}',t)
-f(\vec{r},\vec{v},\delta\vec{r},\delta\vec{v},t)f(\vec{r}-a\hat{\vec{\sigma}},\vec{u},\delta\vec{s},\delta\vec{u},t)]~.
\label{eq:tangboltzmannhard}
\end{eqnarray}
\end{widetext}
If the tangent $\mu$-space variables $\delta \vec{r}$ and $\delta\vec{v}$ are integrated over, this equation reduces to Eq.~(\ref{eq:boltzmannhard}).

Because $\delta\vec{r}$ and $\delta\vec{v}$ are infinitesimal, the dynamics in tangent space are linear in these quantities.
Therefore, from Eqs.~(\ref{eq:collision}) and (\ref{eq:tangboltzmannhard}) one may obtain closed linear equations for the time evolution of the average first moments $\langle\delta\vec{r}\rangle$ and $\langle\delta\vec{v}\rangle$.
To this end, multiply both sides of Eq.~(\ref{eq:tangboltzmannhard}) by the tangent-space vectors and then integrate over them.

The result is a set of equations for the averages,
\begin{eqnarray}
\frac{d}{d t} \delta \vec{r}(\vec{r},\vec{v}, t)
 & = & - \vec{v}\cdot \frac\partial{\partial \vec{r}} \delta\vec{r}(\vec{r},\vec{v}, t)
\nonumber\\
\label{eq:boltzr}
&&\null+ \delta\vec{v}(\vec{r},\vec{v}, t) + {\sf C}_{\sf S}\delta \vec{r}(\vec{r},\vec{v}, t)~,\\
\frac{ d}{ d t} \delta \vec{v} (\vec{r},\vec{v}, t)
& = & - \vec{v}\cdot \frac{\partial}{\partial \vec{r}} \delta \vec{v}(\vec{r},\vec{v}, t)\nonumber\\
&&\null + {\sf C}_{\sf S} \delta \vec{v}(\vec{r},\vec{v}, t) + {\sf C}_{\sf Q} \delta \vec{r}(\vec{r},\vec{v}, t)~.
\label{eq:boltzv}
\end{eqnarray}
The functions $\delta\vec{r}(\vec{r},\vec{v}, t)$ and $\delta\vec{v }(\vec{r},\vec{v}, t)$ are the averages of the tangent $\mu$-space vectors of a  particle, as a function of its position and velocity, and of time.
The linear collision operators  ${\sf C}_{\sf S}$ and ${\sf C}_{\sf Q}$ are  associated with the matrices ${\sf S}$ and ${\sf Q}$, and given by
\begin{widetext}
\begin{eqnarray}
{\sf C}_{\sf S} \, \delta\vec{q} (\vec{r},\vec{v}, t)  = \lefteqn{\int_{\hat{\vec{\sigma}} \cdot (\vec{v}-\vec{u})\leq 0}\, d\vec{u}\, d\hat{\vec{\sigma}} \, \chi_{\mathrm E}(n)n a^{d-1}\, |\,\hat{\vec{\sigma}} \cdot (\vec{v}-\vec{u})|\phi_{\mathrm{M}}(\vec{u})}\nonumber\\
\label{eq:CS}\phantom{\int}&& \times \left\{ \,\delta\vec{q} (\vec{r},\vec{v}', t) + {\sf S} \cdot \left[ \,\delta\vec{q} (\vec{r} + a \hat{\vec{\sigma}},\vec{u}', t)-\,\delta\vec{q} (\vec{r},\vec{v}', t)\right]   -\,\delta\vec{q} (\vec{r},\vec{v}, t) \right\}~,\\
{\sf C}_{\sf Q} \, \delta \vec{r}(\vec{r},\vec{v}, t)  = \lefteqn{ \int_{\hat{\vec{\sigma}} \cdot (\vec{v}-\vec{u})\leq 0}\, d\vec{u}\, d\hat{\vec{\sigma}}
\, \chi_{\mathrm E}(n)n a^{d-1}\, |\,\hat{\vec{\sigma}} \cdot (\vec{v}-\vec{u})|\phi_{\mathrm{M}}(\vec{u})}\nonumber\\
\label{eq:CQ}\phantom{\int}&& \times {\sf Q} \cdot \left[ \delta\vec{r} (\vec{r} + a \hat{\vec{\sigma}},\vec{u}', t) -  \,\delta\vec{r} (\vec{r},\vec{v}', t)\right] ~,
\end{eqnarray}
\end{widetext}
where $\delta\vec{q}$ can be either $\delta\vec{r}$ or  $\delta\vec{v}$.
In Eq.~(\ref{eq:CS}) the first two terms between braces are gain terms.  The last term is the loss term.
Note that, from Eq.~(\ref{eq:Q}), ${\sf Q}$ is a function of the collision parameter and the velocities of the particles before the collision.
This means that in Eq.~(\ref{eq:CQ}) it is a function of $\hat{\sigma}$, $\vec{u}'$, and $\vec{v}'$.
The collision operators are proportional to the collision frequency $\nu$, which for dilute systems is proportional to the number density $n$.

\subsection{\label{sec:fourier}Fourier transform}
As the translation operators commute with the collision operators (\ref{eq:CS}) and (\ref{eq:CQ}),
solutions to Eqs.~(\ref{eq:boltzr}) and (\ref{eq:boltzv}) may be found of the form
\begin{equation}
\delta\vec{q}(\vec{r},\vec{v}, t) = \Delta\vec{q}(\vec{v})\exp({i \vec{k}\cdot\vec{r} + \lambda t})~,
\label{eq:fourier}
\end{equation}
where $\vec{q}$ is either $\vec{r}$ or $\vec{v}$, and $k_j=2\pi n_j/L_j$ is the $j$th component of the wave vector of the sinusoidal modulation and $\lambda$ is the Lyapunov exponent.
Among these the Goldstone modes are those solutions that in the limit of vanishing wave number reduce to 
linear combinations of the zero modes. For these modes to stand out among the continuum of other modes 
their wavelength has to be large compared to the typical length scale of the mean free path, or
\begin{eqnarray}
\label{eq:kklein} k \bar{|\vec{v}|} \ll \nu~.
\end{eqnarray}

On substituting Eq.~(\ref{eq:fourier}) into Eqs.~(\ref{eq:boltzr}) and (\ref{eq:boltzv}), they become eigenvalue equations for the Goldstone modes,
\begin{eqnarray}
\label{eq:boltzrk}\lambda \Delta \vec{r}(\vec{v}) & = & - i (\vec{k}\cdot \vec{v}) \Delta\vec{r} + \Delta\vec{v} + {\sf B}_{\sf S}\Delta \vec{r}~,\\
\label{eq:boltzvk}\lambda \Delta \vec{v}(\vec{v}) & = & - i (\vec{k}\cdot \vec{v}) \Delta \vec{v} + {\sf B}_{\sf S} \Delta \vec{v} + {\sf B}_{\sf Q} \Delta \vec{r}~.
\end{eqnarray}
Spatial propagation, as seen in the simulations for the longitudinal modes, may be accounted for by allowing $\lambda$ to have an imaginary component.
${\sf B}_{\sf S}$ and ${\sf B}_{\sf Q}$ are the Fourier transforms of ${\sf C}_{\sf S}$ and ${\sf C}_{\sf Q}$,
\begin{widetext}
\begin{eqnarray}
{\sf B}_{\sf S}  \Delta\vec{q} (\vec{v})  = \lefteqn{ \int_{\hat{\vec{\sigma}} \cdot (\vec{v}-\vec{u})\leq 0}\, d\vec{u}\, d\hat{\vec{\sigma}} \, \chi_{\mathrm E}(n)n a^{d-1}\, |\hat{\vec{\sigma}} \cdot (\vec{v}-\vec{u})|\phi_{\mathrm{M}}(\vec{u})}\nonumber\\
\label{eq:BS}\phantom{\int}&& \times\left\{ \Delta\vec{q} (\vec{v}') + {\sf S} \cdot \left[ \Delta\vec{q} (\vec{u}') \exp{(-i a \vec{k} \cdot \hat{\vec{\sigma}})}-\Delta\vec{q} (\vec{v}')\right]   -\Delta\vec{q} (\vec{v}) \right\}~,\\
{\sf B}_{\sf Q}  \Delta \vec{r}(\vec{v})  = \lefteqn{ \int_{\hat{\vec{\sigma}} \cdot (\vec{v}-\vec{u})\leq 0}\, d\vec{u}\, d\hat{\vec{\sigma}} \, \chi_{\mathrm E}(n)n a^{d-1} |\hat{\vec{\sigma}} \cdot (\vec{v}-\vec{u})|\phi_{\mathrm{M}}(\vec{u})} \nonumber\\
\label{eq:BQ} \phantom{\int}&& \times {\sf Q} \cdot \left[ \Delta\vec{r} (\vec{u}') \exp{(-i a \vec{k} \cdot \hat{\vec{\sigma}})}-  \Delta\vec{r} (\vec{v}')\right] ~,
\end{eqnarray}
\end{widetext}
where $\Delta\vec{q}$ can be either $\Delta\vec{r}$ or $\Delta\vec{v}$.
One of these, for instance $\Delta \vec{v}$, can be eliminated from the equations.
One must solve Eq.~(\ref{eq:boltzrk}) for $\Delta \vec{v}$ and substitute the result into Eq.~(\ref{eq:boltzvk}).
This yields
\begin{eqnarray}
\label{eq:hoofdvergl}[\left(\lambda + i \vec{k}\cdot \vec{v} - {\sf B}_{\sf S}\right)^2 -{\sf B}_{\sf Q}] \Delta\vec{r} = 0~.
\end{eqnarray}
This equation can be solved by the use of a perturbation expansion in powers of $k$, provided
the mean free path is much smaller than the wavelength, as expressed by Eq.~(\ref{eq:kklein}).
This is done in Sec.~\ref{sec:perturbation}.

\section{\label{sec:solutions}Solutions}
\subsection{\label{sec:perturbation}Perturbation theory}
When Eq.~(\ref{eq:kklein}) is substituted, one may expand the operators and solutions as
\begin{eqnarray}
\label{eq:BSexpansie}{\sf B}_{\sf S}&=& {\sf B}_{\sf S}^{(0)}+k{\sf B}_{\sf S}^{(1)}+k^2{\sf B}_{\sf S}^{(2)}+\dots~,\\
\label{eq:BQexpansie}{\sf B}_{\sf Q}&=& {\sf B}_{\sf Q}^{(0)}+k{\sf B}_{\sf Q}^{(1)}+k^2{\sf B}_{\sf Q}^{(2)}+\dots~,\\
\Delta\vec{r}&=& \Delta\vec{r}^{(0)}+k \Delta\vec{r}^{(1)}+k^2 \Delta\vec{r}^{(2)}+\dots~,
\end{eqnarray}

Note that for $k \rightarrow 0$, the linear operators ${\sf B}_{\sf S}$ and ${\sf B}_{\sf Q}$ become identical to ${\sf C}_{\sf S}$ and ${\sf C}_{\sf Q}$.
When acting on linear combinations of zero modes $\Delta\vec{r}^{(0)}$ they satisfy the properties
\begin{eqnarray}
&\label{eq:0op0modes}{\sf B}_{\sf S}^{(0)}\Delta\vec{r}^{(0)} = {\sf B}_{\sf Q}^{(0)}\Delta\vec{r}^{(0)}\nonumber\\
&= \langle\Delta\vec{r}^{(0)}| {\sf B}_{\sf S}^{(0)}= \langle\Delta\vec{r}^{(0)} |{\sf B}_{\sf Q}^{(0)} = 0~,
\end{eqnarray}
where $\langle.|.\rangle$ represents the inner product defined by integration against a Maxwell distribution of the velocity, which
is the equilibrium distribution.
As it turns out, ${\sf B}_{\sf Q}^{(0)}$ has some nontrivial right eigenfunctions with zero eigenvalues, which will have an important effect on the limiting values of Lyapunov exponents in the limit of 
vanishing density. 
An example of such an eigenfunction is $\Delta\vec{r}(\vec{v})= v_{\perp}\hat{\vec{k}}+v_{\parallel}\hat{\vec{k}}_{\perp}$, where $v_{\perp}$ and $v_{\parallel}$ are the components of $\vec{v}$ perpendicular and parallel to $\vec{k}$.

In zeroth order Eq.~(\ref{eq:hoofdvergl}) reduces to
\begin{eqnarray}
[(\lambda^{(0)} - {\sf B}_{\sf S}^{(0)})^2 - {\sf B}_{\sf Q}^{(0)}]\Delta\vec{r}^{(0)} = 0~.
\end{eqnarray}
The relevant solutions to this are the zero modes, with $\lambda^{(0)}=0$.
This means that the Goldstone modes to leading order in $k$ are the zero modes with a sinusoidal modulation, in nice agreement with the findings
in Refs.~\cite{posch1,posch2}.

In linear order, one finds with the aid of Eq.~(\ref{eq:0op0modes})
\begin{eqnarray}
\label{eq:deltar1}[ -{\sf B}_{\sf S}^{(0)}(i \hat{\vec{k}}\cdot \vec{v} - {\sf B}_{\sf S}^{(1)} ) -{\sf B}_{\sf Q}^{(1)}] \Delta\vec{r}^{(0)}\nonumber\\
 = - [({\sf B}_{\sf S}^{(0)})^2 - {\sf B}_{\sf Q}^{(0)}] \Delta\vec{r}^{(1)}~,
\end{eqnarray}
where $\hat{\vec{k}}$ is the unit vector in the direction of $\vec{k}$.
This is an equation for $\Delta\vec{r}^{(1)}$.
Its formal solution is
\begin{eqnarray}
\Delta\vec{r}^{(1)}&=&
 [ ( {\sf B}_{\sf S}^{(0)} )^2 - {\sf B}_{\sf Q}^{(0)} ]^{-1}\nonumber\\
&&[ {\sf B}_{\sf S}^{(0)}(i \hat{\vec{k}}\cdot \vec{v} - {\sf B}_{\sf S}^{(1)} ) +{\sf B}_{\sf Q}^{(1)}] \Delta\vec{r}^{(0)}~.
\label{eq:deltar11}
\end{eqnarray}
This form
suggests that $\Delta\vec{r}^{(1)}$ is of zeroth order in $n$, just as $\Delta\vec{r}^{(0)}$, but this is actually not the case, because the operator $[({\sf B}_{\sf S}^{(0)})^2 - {\sf B}_{\sf Q}^{(0)}]^{-1}$ acts on functions with  
non vanishing components along the
 nontrivial right zero eigenfunctions of ${\sf B}_{\sf Q}^{(0)}$. This yields
contributions to $\Delta\vec{r}^{(1)}$ of order $1/n$.
One might wonder whether this could cause a divergence in the limit of vanishing density, but that is not the case
because of the restriction imposed on $k$ by Eq.~(\ref{eq:kklein}).

The second order equation involves the first-order Lyapunov exponent $\lambda^{(1)}$, the second-order Lyapunov exponent $\lambda^{(2)}$, and the second-order tangent-space vector $\Delta\vec{r}^{(2)}$,
\begin{eqnarray}
\nonumber &&[\{-{\sf B}_{\sf S}^{(0)},\lambda^{(2)} - {\sf B}_{\sf S}^{(2)}\}_+\nonumber\\
&&\phantom{[+} + (\lambda^{(1)} + i \hat{\vec{k}}\cdot \vec{v} - {\sf B}_{\sf S}^{(1)})^2 -{\sf B}_{\sf Q}^{(2)}] \Delta\vec{r}^{(0)}\nonumber\\
&&+{[ \{-{\sf B}_{\sf S}^{(0)},i \hat{\vec{k}}\cdot \vec{v} - {\sf B}_{\sf S}^{(1)} \}_+ -{\sf B}_{\sf Q}^{(1)}] \Delta\vec{r}^{(1)}}\nonumber\\
&&+[({\sf B}_{\sf S}^{(0)})^2 -{\sf B}_{\sf Q}^{(0)} ]\Delta\vec{r}^{(2)} = 0~,
\end{eqnarray}
where $\{.,.\}_+$ is used to denote the anticommutator of two operators.
On taking the inner product with $\Delta\vec{r}^{(0)}$, all terms involving $\Delta\vec{r}^{(2)}$ vanish as a consequence of Eq.~(\ref{eq:0op0modes}).
The resulting set of equations reads
\begin{eqnarray}
&&\langle\Delta\vec{r}^{(0)} |[(\lambda^{(1)} + i \hat{\vec{k}}\cdot \vec{v} - {\sf B}_{\sf S}^{(1)})^2  - {\sf B}_{\sf Q}^{(2)}]|\Delta\vec{r}^{(0)}\rangle  \nonumber\\
&&\null+ \langle\Delta\vec{r}^{(0)} | [-(i \hat{\vec{k}}\cdot \vec{v} - {\sf B}_{\sf S}^{(1)}) {\sf B}_{\sf S}^{(0)} -{\sf B}_{\sf Q}^{(1)}] | \Delta\vec{r}^{(1)}\rangle \nonumber\\
&&\phantom{nnnn}= 0~.
\label{eq:lambda}
\end{eqnarray}
Since $\Delta\vec{r}^{(0)}$ is a linear combination of three independent zero modes, 
Eq.~(\ref{eq:lambda}) actually has to be read as a $3\times3$ matrix equation involving the matrix 
elements between the various zero modes. In principle all of these are second-order polynomials in 
$\lambda^{(1)}$.
The eigenvalues, as usual, follow from the condition that the determinant of the matrix
vanishes as a function of $\lambda^{(1)}$.

\subsection{General form of the solutions}
In order to investigate the general structure of Eq.~(\ref{eq:lambda}) it is useful to organize
the zero modes for $\Delta\vec{r}^{(0)}$ as
\begin{eqnarray}
\Delta\vec{r}^{(0)}_\perp=\hat{\vec{k}}_{\perp};\quad \Delta\vec{r}^{(0)}_\parallel=\hat{\vec{k}};\quad \Delta\vec{r}^{(0)}_{\vec{v}}=\sqrt{\frac{\beta m}{2}}\,\vec{v}~,
\label{eq:base}
\end{eqnarray}
where $\beta=1/({k_{\mathrm B}T})$.
The first mode consists of a perpendicular displacement, i.e., a spatial translation normal to the 
wave vector, the second mode to a parallel displacement, and the third one to a time translation.

The first mode is odd in $\hat{\vec{k}}_{\perp}$ and the last two even; the first two modes are even in 
$\vec{v}$ and the last one odd. The collision operators ${\sf B}_{\sf S}$ and ${\sf B}_{\sf Q}$ as well 
as the function $\hat{\vec{k}}\cdot \vec{v}$ are odd in $\hat{\vec{k}}_{\perp}$ to every order. The
operators ${\sf B}_{\sf S}^{(n)}$ and ${\sf B}_{\sf Q}^{(n)}$ are even in $\vec{v}$ for even $n$ and odd
for odd $n$. On the basis of these parity properties it follows immediately that the  structure 
of Eq.~(\ref{eq:lambda}), written as a matrix equation on the basis (\ref{eq:base}), is restricted to
\begin{eqnarray}
(\lambda^{(1)})^2 
\left(\begin{array}{lll}
1&0&0\cr
0&1&0\cr
0&0&1\cr
\end{array}\right)
+
i \lambda^{(1)}
\left(\begin{array}{ccc}
0&0&0\cr
0&0&x_{\vec{v},\parallel}\cr
0&x_{\parallel,\vec{v}}&0\cr
\end{array}\right)
\nonumber\\-
\left(\begin{array}{rrr}
y_{\perp,\perp}&0&0\cr
0&y_{\parallel,\parallel}&0\cr
0&0&y_{\vec{v},\vec{v}}\cr
\end{array}\right)
= 0
\label{eq:genform}
~.
\end{eqnarray}
The constants $x$ and $y$ are determined by temperature and by the form of the collision operators.
From this it becomes clear that the equation can be split into two parts, one for the perpendicular zero mode, the transverse part, and one for the parallel zero mode and the time mode, the longitudinal part.
From the general form of the matrices one can derive the general form of the Lyapunov exponents $\lambda = k \lambda^{(1)}$ to be
\begin{eqnarray}
\label{eq:gentrans}\lambda_{\mathrm{trans}} &  =  &\pm k \sqrt {y_{\perp,\perp}}~,\\
\label{eq:genlong}\lambda_\mathrm{long} & = & \pm k \sqrt{ y_1 \pm i\sqrt{y_2}}~,
\end{eqnarray}
where $y_1$ and $y_2$ are functions of $x_{\vec{v},\parallel}, x_{\parallel,\vec{v}}, y_{\parallel,\parallel}$, and $y_{\vec{v},\vec{v}}$.
If $y_{\perp,\perp} > 0$, the Lyapunov exponent of the transverse mode is real and therefore the mode is of the same form as in the simulations reported in \cite{posch1,posch2}.
If $y_2> 0 $, the longitudinal Lyapunov exponents have both real and imaginary components, and these modes also have the form of the longitudinal modes found in the simulations.

\subsection{Density expansion \label{sec:density}}
The Sto{\ss}zahlansatz is an approximation that for many purposes, e.g., the derivation of hydrodynamic equations from the 
Boltzmann equation with explicit expressions for the transport coefficients~\cite{chapmancowling}, 
becomes exact in the limit of vanishing density. Therefore we want to investigate the behavior of our 
equations in this limit and compare to the results found in the simulations.
In the limit of vanishing density Eq.~(\ref{eq:lambda}) becomes
\begin{eqnarray}
&&
\langle\Delta\vec{r}^{(0)} |[\lambda^{(1)} + i (\hat{\vec{k}}\cdot \vec{v})]^2|\Delta \vec{r}^{(0)}\rangle \nonumber\\
&&\null- \langle\Delta\vec{r}^{(0)} | [ i (\hat{\vec{k}}\cdot \vec{v}) {\sf B}_{\sf S}^{(0)} +{\sf B}_{\sf Q}^{(1)} ]| \Delta\vec{r}^{(1)}\rangle = 0~.
\label{eq:lambdan0}
\end{eqnarray}
For this equation it is crucial indeed 
that $\Delta \vec{r}^{(1)}$ is of the order of $n^{-1}$.
If there were no nontrivial right eigenfunctions of ${\sf B}_{\sf Q}^{(0)}$ with eigenvalue 0, $\Delta\vec{r}^{(1)}$ would be one order of $n$ higher, and the second term would not contribute to the Lyapunov exponents in the limit of vanishing density.

In the following section we will further discuss the actual magnitudes of the two terms in 
Eq.~(\ref{eq:lambdan0}).

\section{Results and discussion\label{sec:results}}
If only the first term in Eq.~(\ref{eq:lambda}) is kept, the calculation is fairly simple.
From now on we choose $d=2$.
The same calculations can easily be repeated for $d=3$, but there are far fewer simulation results to compare to.
Equation (\ref{eq:genform}) in this approximation becomes
\begin{eqnarray}
(\lambda^{(1)})^2{\frac{\beta m}{2}}
\left(\begin{array}{lll}
1&0&0\cr
0&1&0\cr
0&0&1\cr
\end{array}\right)
+
i \lambda^{(1)} \sqrt{\frac{\beta m}{2}}
\left(\begin{array}{ccc}
0&0&0\cr
0&0&1\cr
0&1&0\cr
\end{array}\right)
\nonumber\\-
\left(\begin{array}{rrr}
\frac{1}{2}&0&0\cr
0&\frac{1}{2}&0\cr
0&0&1\cr
\end{array}\right)
= 0
~,
\end{eqnarray}
independent of the density.
The solutions for the Lyapunov exponents then are
\begin{eqnarray}
\label{eq:geendeltar1begin}
\lambda_{\mathrm{trans}} &=& \pm \frac{k}{\sqrt{\beta m}} ~,\\
\lambda_{\mathrm{long}} &=& \pm \frac12 \sqrt{1\pm i\sqrt{7}}  \sqrt{\frac{2}{\beta m}} k\\
&\approx& (\pm 0.978 \pm i~0.676) \frac{k}{\sqrt{\beta m}}~.
\label{eq:geendeltar1eind}
\end{eqnarray}
The structure of the corresponding eigenvectors is indeed like that found in simulations \cite{posch1,posch2}.

To calculate the contribution from the second term in Eq.~(\ref{eq:lambda}) to the leading order of the Lyapunov exponents, one has to choose a suitable basis in which to express the function $\Delta\vec{r}^{(1)}(\vec{v})$.
The basis must be orthogonal with regard to the chosen inner product $\langle . | . \rangle$.
Next, the matrix elements of the operators ${\sf B}_{\sf S}^{(i)}$ and ${\sf B}_{\sf Q}^{(i)}$ must be calculated between elements of the basis.

A simple, but suitable, basis is the set of functions that are products of Hermite polynomials in the components of $\vec{v}\sqrt{m \beta/2}$ parallel and perpendicular to the wave vector $\vec k$.
The Hermite polynomials $H_i(x)$ form a complete orthogonal basis with regard to integration against $\exp{(-x^2)}$, and therefore their products will be orthogonal under the inner product used here.
The solution to Eq.~(\ref{eq:hoofdvergl}) can thus be expanded as
\begin{eqnarray}
\Delta\vec{r}(\vec{v}) = \sum_{l,p,q} \vec{e}_l\, c_{l,p,q} H_p(v_\parallel) H_q(v_\perp)~,\label{eq:hermitebasis}
\end{eqnarray}
where $l$ can be either $\perp$ or $\parallel$, $\vec{e}_\perp$ is $\hat{\vec{k}}_\perp$ and $\vec{e}_\parallel$ is $\hat{\vec{k}}$.

By truncating all expressions at some finite order in the polynomial expansion, one finds approximate values for $\lambda^{(1)}$.
For good convergence one has to go beyond the zeroth- and first-order Hermite polynomials.
In the Appendix more details are given on the matrix representations of the truncated operators  and on the convergence of the Lyapunov exponents in dependence on the order of truncation.

\begin{figure}
\includegraphics[height=8.6cm,angle=270]{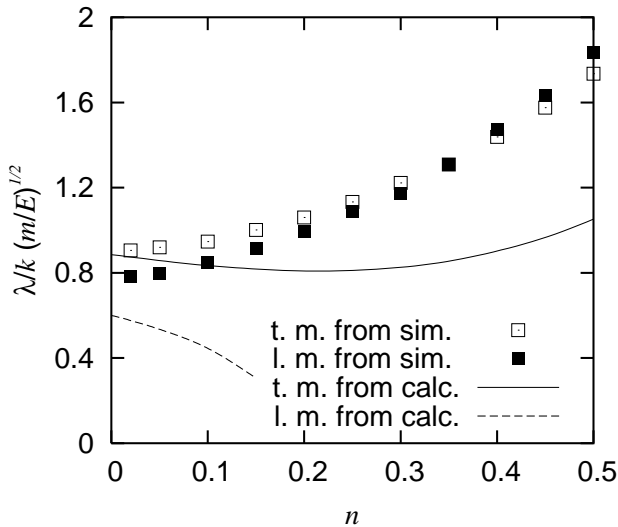}
\caption{\label{fig:exponentjes}The Lyapunov exponents for transverse and longitudinal modes from the simulations \cite{posch1,posch2} compared to the present calculations.}
\end{figure}
\begin{figure}
\includegraphics[height=8.6cm,angle=270]{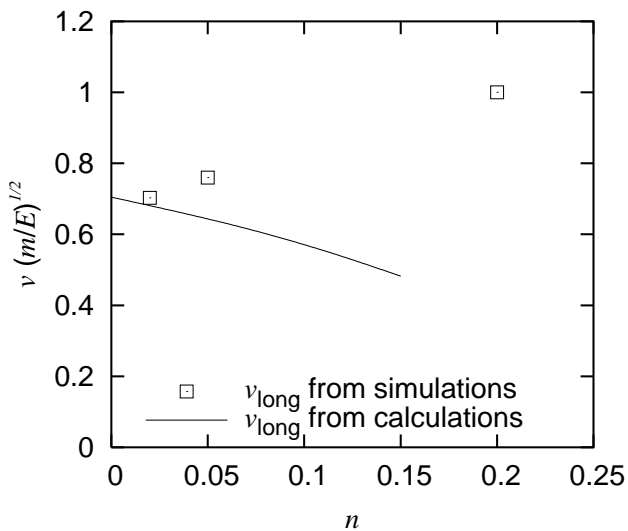}
\caption{\label{fig:vel} The  velocities of the longitudinal mode from simulations \cite{posch1,posch2} compared to the present calculations.}
\end{figure}

To sixth order in the polynomial expansion, the results for $n\rightarrow 0$ and $d=2$ are
\begin{eqnarray}
\label{eq:resultst}\lambda_{\mathrm{trans}} =& \pm 0.886 &\frac{k}{\sqrt{\beta m}} ~,\\
\label{eq:resultsl}\lambda_{\mathrm{long}} =& \pm 0.607 &\frac{k}{\sqrt{\beta m}}~,\\
\label{eq:resultsv}v_{\mathrm{long}}  =&  \pm 0.706 &\frac{1}{\sqrt{\beta m}}~.
\end{eqnarray}
With respect to Eqs.~(\ref{eq:geendeltar1begin})-(\ref{eq:geendeltar1eind}) the corrections are largest for the Lyapunov exponent of the longitudinal mode.

For low densities the form of the modes is predicted correctly by the calculations; the modes are split into non propagating transverse and propagating longitudinal modes.
For number density $n=\rho a^2=0.02$, the Lyapunov exponents from the simulations are
\begin{eqnarray}
\label{eq:lambdat} {\lambda_{\mathrm{trans}}}  = & \pm 0.906 &\frac{k}{\sqrt{\beta m}}~,\\
\label{eq:lambdal} {\lambda_{\mathrm{long}}}  = & \pm 0.783 &\frac{k}{\sqrt{\beta m}}~,\\
\label{eq:vl} v_{\mathrm{long}}  = & \pm 0.703&\frac{1}{\sqrt{\beta m}}~.
\end{eqnarray}
The calculated Lyapunov exponent of the transverse mode and the propagation speed of the longitudinal mode compare to the values from the simulations, within 2\% at $n=0.02$.
The Lyapunov exponent for the longitudinal mode deviates by about 30\%.

The results for higher densities are displayed in Fig.~\ref{fig:exponentjes}.
With increasing density the calculated values deviate increasingly from the simulation results.
For the longitudinal mode the predicted real part of the Lyapunov exponent even drops to 0 and the exponent becomes purely imaginary.

The deviations from the simulations can be attributed to contributions from ring terms and 
possibly other contributions to a generalized ${\sf B}_{\sf Q}$ operator that are at most of order 
$n^2$. From Eq.~(\ref{eq:deltar11}) one sees that such terms, working on the non trivial zero 
eigenfunctions of ${\sf B}_{\sf Q}^{(0)}$, contribute to the leading order terms in the density 
expansion of $\Delta\vec{r}^{(1)}$, just like $({\sf B}_{\sf S}^{(0)})^2$. Therefore they have to be 
included in the  second term in Eq.~(\ref{eq:lambdan0}).
So in contrast to usual applications of kinetic theory, where ring terms only contribute to higher orders in the density, in the present case they contribute to the leading order.
In the present calculation these contributions
are not included, but we are actively working on their evaluation.
Similarly, the ring terms will contribute to higher orders in a density expansion of the Lyapunov exponents.
These contributions may be responsible for the discrepancies between simulation results and Enskog theory for higher densities, which show up in Figs.~\ref{fig:exponentjes} and \ref{fig:vel}.
For more details on ring terms in kinetic theory, see Ref.~\cite{sengers}.

It is interesting to compare our results to those by McNamara and Mareschal \cite{mareschal}, who also based their work on kinetic theory calculations.
They do not derive equations for the distribution functions, but go directly to hydrodynamic like equations for the moments.
To close these, they make hypotheses to factorize the fluxes.
The resulting values for the Lyapunov exponents in the low-density limit are less close to the simulation values than those from our calculations.
It is not clear that in this treatment the effects of the non trivial zero eigenfunctions of ${\sf B}_{\sf Q}^{(0)}$ are accounted for.

Forster and Posch have also done simulations on similar systems with soft potentials \cite{soft}. They
roughly find a branch again of Lyapunov exponents close to zero, but the sinuoidal structure of the 
corresponding modes is much less clear \cite{soft}.
It would be very interesting to calculate the Lyapunov exponents with kinetic theory methods also for this case.
It would also be very interesting to look at small Lyapunov exponents in non equilibrium systems.
However, in such systems the calculations become more complicated because the stationary velocity distributions are not Maxwellian any more.

\section{Conclusion}
In this paper, we have demonstrated how Lyapunov exponents close to zero can be related to Goldstone modes.
We found the correct types of behavior in dependence on the wave number  of the exponents and their tangent-space eigenvectors.
This was achieved through a kinetic theory approach, in which we used a molecular chaos assumption for the pair distribution function to derive an equation similar to the Enskog equation.
For low densities this reduces effectively to a generalized Boltzmann equation.

The calculated values for the exponents belonging to the transverse modes at low densities are within a few percent of the values found in the simulations \cite{posch1,posch2}.
The propagation velocity for the transverse mode is within 1\% of the simulation values.
The value for the Lyapunov exponent belonging to the longitudinal mode deviates from the simulations by 30\%.
For higher densities, the predicted values deviate increasingly more from the values
found in the simulations.
These deviations are probably due to contributions from ring collisions and similar terms.
In most applications of the Boltzmann equation and the Sto{\ss}zahlansatz such terms produce contributions to the relevant quantities which are one order of higher order in the density, but in the problem at hand they turn out to contribute to the leading order.

\begin{acknowledgments}
We wish to thank H.~A.~Posch, R.~Hirschl, and C.~Forster for kindly providing us with their simulation program.
\end{acknowledgments}

\appendix
\section{Expansion in Hermite Polynomials}

In order to solve Eq.~(\ref{eq:lambda}), one must write the operators in Eqs.~(\ref{eq:BS}) and (\ref{eq:BQ}) as matrices between the basis functions described in Eq.~(\ref{eq:hermitebasis}).
From now on we take $d=2$, but the same calculations can easily be done for three dimensions.
We only show results for basis functions of up to linear order in $\vec{v}$.  In Eq.~(\ref{eq:hermitebasis}) $p$ and $q$ can be equal to 0 or 1.
In fact one has to include higher powers to find good approximations for the solutions to the original equations.
If the first component is the component parallel to $\vec{k}$, the basis is ordered as $(1,0)$; $(0,1)$; $(\sqrt{\beta m/2} \,v_\parallel,0)$; $(0,\sqrt{\beta m/2}\, v_\parallel)$; $(\sqrt{\beta m/2}\, v_\perp,0)$; $(0,\sqrt{\beta m/2}\, v_\perp)$.
All coefficients are given to leading order in $n$.
In this notation the zero modes are 
\begin{eqnarray}
\Delta \vec{r}^{(0)}_1 &=& (1,0,0,0,0,0)~,\\
&&(0,1,0,0,0,0)~,\\
&&(0,0,{\textstyle \frac12} \sqrt 2,0,0,{\textstyle \frac12} \sqrt 2)~.
\end{eqnarray}
Here, the subscript 1 indicates that basis functions up to first order in $\vec{v}$ have been included.
From Eqs.~(\ref{eq:deltar1}) and (\ref{eq:lambda}) it follows that the operator ${\sf B}_{\sf S}$ described in Eq.~(\ref{eq:BS}) is only needed up to first order in $k$.
One finds for the matrix elements of ${\sf B}_{\sf S}$ in the expansion of Eq.~(\ref{eq:BSexpansie})
\begin{align}
{\sf B}_{\sf S,1}^{(0)} =& \sqrt{\frac{2}{\beta m}}\, \frac{3 \sqrt{2 \pi}}8\,  n a&
\left(\begin{array}{rrrrrr}
0&0&0&0&0&0\cr
0&0&0&0&0&0\cr
0&0&\llap{-}1&0&0&1\cr
0&0&0&\llap{-}3&1&0\cr
0&0&0&1&\llap{-}3&0\cr
0&0&1&0&0&\llap{-}1
\end{array}\right)~, \displaybreak[1] \\
{\sf B}_{\sf S,1}^{(1)} = & \sqrt{\frac{2}{m \beta}}\, \frac{\sqrt2\,\pi i}{16}\, n a^2&  
\left(\begin{array}{rrrrrr}
0&0&3&0&0&1\cr
0&0&0&1&1&0\cr
3&0&0&0&0&0\cr
0&1&0&0&0&0\cr
0&1&0&0&0&0\cr
1&0&0&0&0&0
\end{array}\right)~.
\end{align}
The first three contributions to ${\sf B}_{\sf Q}$ as expanded in Eqs.~(\ref{eq:BQ}) and (\ref{eq:BQexpansie})
have similar matrix representations of the form
\begin{align}
{\sf B}_{\sf Q,1}^{(0)} = & {\frac{2 }{\beta m}}\,  2 \pi\, n & 
\left(\begin{array}{rrrrrr}
0&0&0&0&0&0\cr
0&0&0&0&0&0\cr
0&0&0&0&0&0\cr
0&0&0&\llap{-}1&1&0\cr
0&0&0&1&\llap{-}1&0\cr
0&0&0&0&0&0
\end{array}\right)~, \\\displaybreak[1]
{\sf B}_{\sf Q,1}^{(1)} = & {\frac{2}{\beta m}}\, \frac{\sqrt\pi\, i}8\, na& 
\left(\begin{array}{rrrrrr}
0&0&1&0&0&\llap{-}1\cr
0&0&0&\llap{-}5&7&0\cr
1&0&0&0&0&0\cr
0&\llap{-}5&0&0&0&0\cr
0&7&0&0&0&0\cr
\llap{-}1&0&0&0&0&0
\end{array}\right)~, \\\displaybreak[1]
{\sf B}_{\sf Q,1}^{(2)} = & {\frac{2}{\beta m}}\, \frac\pi8\, n a^2& 
\left(\begin{array}{rrrrrr}
\llap{-}2&0&0&0&0&0\cr
0&2&0&0&0&0\cr
0&0&\llap{-}1&0&0&0\cr
0&0&0&5&\llap{-}4&0\cr
0&0&0&\llap{-}4&3&0\cr
0&0&0&0&0&1
\end{array}\right)~.
\end{align}
The operators $i \hat{\vec{k}}\cdot \vec{v}$ and $-(\hat{\vec{k}}\cdot\vec{v})^2$ can also be written in this way.
One finds
\begin{align}
i \hat{\vec{k}}\cdot \vec{v} = & - \sqrt{\frac{1 }{\beta m}} &
\left(\begin{array}{rrrrrr}
0&0&i&0&0&0\cr
0&0&0&i&0&0\cr
i&0&0&0&0&0\cr
0&i&0&0&0&0\cr
0&0&0&0&0&0\cr
0&0&0&0&0&0
\end{array}\right)~, \\\displaybreak[1]
-(\hat{\vec{k}}\cdot\vec{v})^2 = & - \frac12 \,\sqrt{\frac{2}{\beta m}} &
\left(\begin{array}{rrrrrr}
1&0&0&0&0&0\cr
0&1&0&0&0&0\cr
0&0&3&0&0&0\cr
0&0&0&3&0&0\cr
0&0&0&0&1&0\cr
0&0&0&0&0&1
\end{array}\right)~.
\end{align}

With these matrices and Eq.~(\ref{eq:deltar1}) the vectors for $\Delta \vec{r}^{(1)}$ may be expressed in terms of $\Delta\vec{r}^{(0)}$ up to first order in the polynomial expansion in $\vec{v}$.
With the orthogonality relation between $\Delta\vec{r}^{(1)}$ and $\Delta\vec{r}^{(0)}$, mentioned in Sec.~\ref{sec:perturbation}, this yields
\begin{equation}
\Delta\vec{r}^{(1)}_1 = -\frac{2 i}{9 n a \sqrt{\pi}} 
\left(\begin{array}{rrrrrr}
0&0&0&0&0&0\cr
0&0&0&0&0&0\cr
1&0&0&0&0&0\cr
0&1&0&0&0&0\cr
0&1&0&0&0&0\cr
\llap{-}1&0&0&0&0&0
\end{array}\right)
\cdot \Delta\vec{r}^{(0)}_1~.
\end{equation}
These matrices can be used to find the $3 \times 3$ matrices in Eqs.~(\ref{eq:lambda}) and (\ref{eq:genform}).
The equation to leading order in $n$ then becomes
\begin{widetext}
\begin{equation}
\det
\left(
\begin{array}{ccc}
-\frac{7}{18}+ \left(\lambda^{(1)}_1 \sqrt{\frac{\beta m}{2}} \right)^2&0&0\cr
0&-\frac{7}{18}+ \left(\lambda^{(1)}_1 \sqrt{\frac{\beta m}{2}}\right)^2&i\lambda^{(1)}_1  \sqrt{\frac{\beta m}{2}}\cr
0&i\lambda^{(1)}_1  \sqrt{\frac{\beta m}{2}}&-1+ \left(\lambda^{(1)}_1  \sqrt{\frac{\beta m}{2}}\right)^2\cr
\end{array}
\right) = 0~.
\end{equation}
\end{widetext}
Here, the indices of the matrix on the left hand side are ordered according to $\Delta\vec{r}^{(0)}_{\perp} , \Delta\vec{r}^{(0)}_{\parallel}, \Delta\vec{r}^{(0)}_{\vec{v}}$.
The matrix can be factorized into two parts.
One part describes  the transverse mode and produces a simple quadratic equation for $\lambda^{(1)}$, with the solution
\begin{eqnarray}
\lambda^{(1)}_1 \sqrt{\frac{\beta m}{2}} = \pm {\frac16} \sqrt{14}\approx \pm 0.62361~.
\end{eqnarray}
The remaining  part yields the longitudinal mode.
The solutions to the fourth order equation for $\lambda^{(1)}$ are
\begin{eqnarray}
\lambda^{(1)}_1 \sqrt{\frac{\beta m}{2}} &=& \pm \frac16 \sqrt{ (7 \pm i \sqrt{455})}
\nonumber\\
&\approx& \pm 0.639552 \pm 0.493231~i~.
\end{eqnarray}
The same calculation can be done with larger subsets of the basis.
The results are shown in Table~\ref{tab:tabelletje}.

\begin{table}
\begin{ruledtabular}
\begin{tabular}{r|l|l}
$n$&$\mathrm{transverse~} \lambda^{(1)}_n\sqrt{\frac{\beta m}{2}} $&$\mathrm{longitudinal~} \lambda^{(1)}_n\sqrt{\frac{\beta m}{2}}$\\
\hline
1&$\pm0.62361$&$\pm 0.639552\pm 0.463231~i$\\
2&$\pm0.62361$&$\pm 0.422807\pm 0.499026~i$\\
3&$\pm0.626194$&$\pm 0.424806 \pm 0.499105~i$\\
4&$\pm0.626194$&$\pm 0.428599 \pm 0.498952~i$\\
5&$\pm 0.626254$&$\pm 0.428645 \pm 0.498954~i$\\
6&$\pm 0.626254$&$\pm 0.429104 \pm 0.498953~i$
\end{tabular}
\end{ruledtabular}
\caption{\label{tab:tabelletje} The Lyapunov exponents and the propagation velocities for the longitudinal mode calculated using products of Hermite polynomials in $\vec{v}$ up to different orders.}
\end{table}

Using functions up to an odd power in $\vec{v}$ is different from using functions up to an even power, because the odd powered functions contribute  to different matrix elements than the even powered functions.
To determine whether the solutions have converged, one must therefore look at the behavior as the maximum power is increased by steps of 2.
The error in the results using up to sixth powers in $\vec{v}$ can be estimated by comparing the values with the results for powers in $\vec{v}$ up to four.
The error in the solutions when using up to sixth powers of $\vec{v}$ in the basis functions appears not to be much larger than a tenth of a percent, except in the case of the longitudinal mode, where it might be of the order of a few tenth of a percent.

\end{document}